\documentclass[superscriptaddress,pre,
               preprintnumbers,amsmath,amssymb]{article} 
\pdfoutput=1

\usepackage{color}
\usepackage{graphicx}
\usepackage{setspace}
\usepackage[round]{natbib}

\newcommand{\dint}{\displaystyle\int}

\newcommand{\rr}[1]{{\normalfont\textrm{#1}}}
\newcommand{\cc}[1]{{\mathcal{#1}}}
\definecolor{light}{gray}{.9}

\title{Phase transition in saturated porous media: pore--fluid segregation in consolidation}

\author{Emilio N.M.\ Cirillo$^1$, Nicoletta Ianiro$^1$ and Giulio Sciarra$^{2,}$\footnote{Corresponding author: e-mail giulio.sciarra@uniroma1.it Tel. +390644585230 Fax +390644585618}\medskip
\and \centering $^1$ Dipartimento Me.\ Mo.\ Mat., Universit\`a di Roma \lq\lq La Sapienza\rq\rq,
\and \centering via A.\ Scarpa 16, 00161 Rome, Italy\medskip
\and \centering $^2$ Dipartimento Ingegneria Chimica Materiali Ambiente,
\and \centering Universit\`a di Roma \lq\lq La Sapienza\rq\rq, via Eudossiana 18, 00184 Rome, Italy}
\begin{document}
\bibliographystyle{plainnat}
\maketitle

\begin{abstract}
\noindent Consider the consolidation process typical of soils, this phenomenon is expected not to exhibit a unique state of equilibrium, depending on the \textit{external loading} and the constitutive parameters. Beyond the standard solution, also pore--fluid segregation, which is typically associated with fluidization of the granular material, can arise. Pore--fluid segregation has been recognized as a phenomenon typical of the short time behavior of a saturated porous slab or a saturated porous sphere, during consolidation. In both circumstances Biot's three dimensional model provides time increasing values of the water pressure (and fluid mass density) at the center of the slab (or of the sphere), at early times, if the Lam\'{e} constant $\mu$ of the skeleton is different from zero. This localized pore--fluid segregation is known in the literature as Mandel--Cryer effect. 
In this paper a non linear poromechanical model is formulated. The model is able to describe the occurrence of two states of equilibrium and the switching from one to the other by considering a kind of \textit{phase transition}. Extending classical Biot's theory a more than quadratic strain energy potential is postulated, depending on the strain of the porous material and the variation of the fluid mass density (measured with respect to the skeleton reference volume). When the consolidating pressure is strong enough the existence of two distinct minima is proven. 
\end{abstract}
\vskip .5cm


%%%%%%%% Il lavoro
%\newpage
\section{I\lowercase{ntroduction}}
\label{s:intro}
\par\noindent
Porous media confined into a fluid infinite reservoir, or suffering consolidation loading, can exhibit solid--solid and solid--fluid phase transitions. These last can be observed because of different phenomena, as gravity driven solid--fluid separation of suspensions \citep[see][]{Burger00} or solid--fluid segregation of soils \citep[see e.g.][]{Nichols94, Vardoulakis04_1, Vardoulakis04_2}. It will be the purpose of this paper to investigate this last transition.

The \citet{biot41, biot55} three dimensional linear model describes short time pore--fluid segregation, in case of consolidation, only for special geometries of the porous material, a slab \citep{Mandel53} or a sphere \citep{Cryer63}. The existence of a fluid richer stationary state of the porous medium can not be proven in the context of linear Biot theory endowed with Darcy solid--fluid viscous coupling. Recently, in-situ observations and experimental studies have pointed out the formation of compaction bands in rocks and soils \citep{Mollema96}. This phenomenon is typically connected with the occurrence of pore--fluid segregation \citep{Holcomb03, Holcomb07} in consolidation processes and, eventually, soil fluidization \citep{Kolymbas98}: under the effect of an applied external pressure, or because of gravity, solid internal remodeling can induce the formation of non--connected fluid--filled cavities. Thus increasing the external loading causes fluid to remain trapped and therefore the fluid mass in the trapping chambers to increase with respect to that of the fluid flowing out of the solid matrix. The pore--fluid pressure is now capable to induce unbalance of forces acting on the soil grains, so causing fluidization.
\begin{figure}[htp]
 \centering
 \includegraphics[width=9cm]{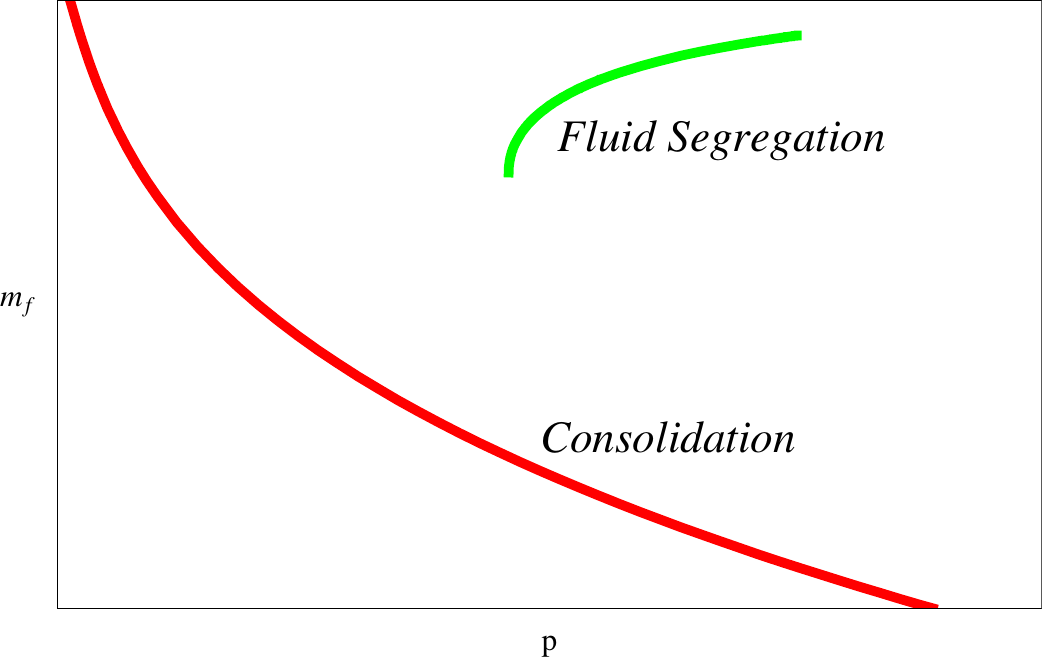}
 % PlotmpIntro.pdf: 300x193 pixel, 72dpi, 10.58x6.81 cm, bb=0 0 300 193
\caption{Qualitative picture of the behavior of the fluid mass, $m_{\rr f}$, with respect to the consolidating pressure $p$.}
 \label{figIntro}
\end{figure}
\citet{Nichols94} have experimentally demostrated the existance of two phases in granular materials saturated by a fluid. A cylindrical perspex test vessel is filled by a granular test sample and water is injected from the bottom through the granular layer. Tuning water pressure increases the flow velocity and therefore the upward drag action on the grains. At low flow velocity, when the drag force is smaller than gravity, the standard phase is observed: the fluid flows through the solid which remains undeformed. When the velocity is increased another phase shows up: the drag force balances gravity and fluidization of the grains occurs. 

The aim of this paper is to model solid--fluid phase transition describing in particular the showing up of a pore--fluid segregating state, when a consolidating external pressure is applied to the porous material. For this pourpose we use a simple one--dimensional model, generalizing the Biot theory, where the phase transition is achieved modifing the standard Biot internal energy functional. A suitable potential energy including the effects of the external pressure $p$ will be considered.

Also in this model the free energy depends on two fields, the deformation
$\varepsilon$ of the porous matrix and the density of the fluid $m_{\rr f}$, measured 
with respect to the solid reference
volume. We shall describe the following phase transition:
there exists $p_\rr{c}>0$ such that for
$0\le p\le p_\rr{c}$, that is \textit{low pressure},
there exists a single stationary state with fluid mass $m_{{\rr f}\,1}(p)$ and 
solid deformation $\varepsilon_1(p)$; while for 
$p>p_\rr{c}$, that is \textit{high pressure},
a second phase $(m_{{\rr f}\,2}(p),\varepsilon_2(p))$ 
appears. Fluid density $m_{{\rr f}\,2}$ is greater than $m_{{\rr f}\,1}$ and increases with the pressure $p$, see Fig \ref{figIntro}.

The first solution corresponds to the case when the fluid is not confined inside the matrix and can flow freely back and forth from the inside to the outside of the solid. At equilibrium the density of the internal fluid equals that of the external infinite reservoir or, in other words the pore of the solid matrix are connected. The second solution, on the other hand, corresponds to the case when the mass density of the fluid is not that of the fluid in the external reservoir. This is due to the fact that the porous material starts to behave as a closed system rather than an open system, as the pore connecting ducts become thinner and thinner.

This point can be supported by means of a thermodynamic argument:
the flow of the liquid back and forth from the inside to the outside of the 
porous medium is a thermodynamic transformation at constant 
temperature, pressure, and volume, typical of open systems. The equilibrium is achieved when the 
internal Gibbs free energy $G$ equals the external one; if the infinitesimal 
mass $\rr{d}m$ of fluid exits the matrix, the mass outside will vary of 
the amount $-\rr{d}m$. 
Since the infinitesimal variation of the internal and 
the external Gibbs free energy are given by 
$\rr{d}G_\rr{i}=\mu_\rr{i}\rr{d}m_\rr{i}=\mu_\rr{i}\rr{d}m$
and
$\rr{d}G_\rr{e}=\mu_\rr{e}\rr{d}m_\rr{e}=-\mu_\rr{e}\rr{d}m$
respectively, 
where $\mu$ is the chemical potential of the fluid,
we have that at equilibrium 
$\rr{d}G_\rr{i}=\rr{d}G_\rr{e}$ implies $\mu_\rr{i}=\mu_\rr{e}$. 
Recalling that $\mu=\partial U/\partial m$, with $U$ the internal energy of the 
fluid, for reasonable choices of the function $U$, we have that the 
equality of the internal and external chemical potential reflects
into the equality of the internal and external fluid density.

It is quite natural that, supposed to limit our discussion to small values of the external pressure, the deformation of the 
solid is grossly proportional to the external pressure $p$, namely, 
$\varepsilon(p)\approx -B\,p$ for some positive 
constant $B$ (recall that 
for a compressed solid matrix the deformation is negative).
In the unique low pressure phase, since as a consequence 
of the pressure $p$ some liquid will exit the solid, we suppose that 
$m_{{\rr f}1}(p)$ decreases proportionally
to $p$. Concerning the second phase $m_{{\rr f}2}(p)$, the only constraint will be 
$m_{{\rr f}2}(p)>m_{{\rr f}1}(p)$. Indeed in this phase we guess the solid structure 
is modified and room for some liquid is made.

In spite of the appealing simplicity of these arguments, it turns out that the actual situation 
is complex to be studied. The first goal is to understand basic physical phenomena
and to determine how phase transition in question is affected by the external pressure. 

The present paper is organized as follows. In \S 2 we describe the generalized
Biot model and introduce the free energy functional. In \S 3, via an analytical minimization of the functional,
we study the stationary points and their character. Finally in \S 4 we discuss our results.

\section{T\lowercase{he model}}
\label{s:modello}
\par\noindent
\textit{Kinematics.\/}
Let $\mathcal B_\rr{s},\, \mathcal B_\rr{f}$ be the \textit{reference}
configurations of the solid and fluid components; and $\mathcal E$ the
Euclidean space of positions.
To specify the \textit{current} configuration of the system, that is the configuration
at time $t\in I\!\! R$, two families of diffeomorphisms, $\{\chi_{\rr{s},t}:\mathcal B_\rr{s}\rightarrow\mathcal E,\,t\in I\!\!R\}$ and 
$\{\phi_{\rr{f},t}:\mathcal B_\rr{s}\rightarrow\mathcal B_\rr{f},\ t\in I\!\!R\}$ are introduced. 

The current solid configuration is given by the \textit{solid placement} map 
$\chi_{\rr{s},t}$; for any $X_\rr{s}\in\mathcal{B}_\rr{s}$, $x=\chi_{\rr{s},t}(X_\rr{s})$ is the 
position occupied, at time $t$ in the Euclidean space $\mathcal E$, by the solid material particle $X_\rr{s}$. The map $\phi_{\rr{f},t}$, on the other hand, identifies the fluid material particle $X_\rr{f}$ in $\mathcal B_\rr{f}$ which, at time $t$, occupies the same current place $x$ as the solid particle $X_\rr{s}$. This description of the kinematics of the fluid is completely consistent with the Eulerean point of view adopted in standard fluid mechanics: the focus is not on the placement of the fluid particles, but on the particle, which at time $t$, occupies the current place $x$. As a consequence the reference configuration of the solid $\mathcal B_\rr{s}$, which in the following will be the \textit{the reference configuration of the system}, is assumed to be a known subdomain of $\mathcal E$, while the one of the fluid, $\mathcal B_\rr{f}$, is unknown, to be determined by the map $\phi_{\rr{f},t}$. As usual also the current configuration of the solid is unknown. Bearing in mind the definition of the map $\phi_{\rr{f},t}$ we shall call
$\mathcal B:=\chi_{\rr{s},t}\left(\mathcal B_{\rr{s}}\right)$ be the \textit{current configuration of the system}. The so called \textit{fluid placement} map $\chi_{\rr{f},t}:\mathcal B_\rr{f}\to\mathcal E$ can be constructed starting from $\chi_{\rr{s},t}$ and $\phi_{\rr{f},t}$. Indeed once we set $\chi_{\rr{f},t}:=\chi_{\rr{s},t}\circ\phi_{\rr{f},t}^{-1}$, for any $X_\rr{f}\in B_\rr{f}$, $\chi_{\rr{f},t}(X_\rr{f})$ represents the position occupied in the current configuration by the fluid particle $X_\rr{f}$. For more details we refer to \citet{Sciarra08}.\medskip

\par\noindent
\textit{Strain.\/}
Let $F_{\rr{s},t}:=\nabla\chi_{\rr{s},t}$ and $\Phi_{\rr{f},t}:=\nabla\phi_{\rr{f},t}$
be the gradients of the maps $\chi_{\rr{s},t}$ and $\phi_{\rr{f},t}$, respectively (to clarify notations we remark that $\nabla$ indicates, in this context, spatial derivative independently of the domain of the map on which it operates). $F_{\rr{s},t}$ is typically named deformation of the solid. Since derivatives are taken with respect to $X_\rr{s}\in \mathcal B_\rr{s}$, those gradients are usually called Lagrangean gradients. We also define the gradient $F_{\rr{f},t}:=\nabla\chi_{\rr{f},t}$;
the chain rule yields
$F_{\rr{f},t}(X_\rr{f})
 =\nabla\chi_{\rr{f},t}(X_\rr{f})
 =\nabla(\chi_{\rr{s},t}(\phi_{\rr{f},t}^{-1}(X_{\rr{f}})))
 =F_{\rr{s},t}(\phi_{\rr{f},t}^{-1} (X_\rr{f}))\nabla\phi_{\rr{f},t}^{-1}(X_{\rr{f}})
 =F_{\rr{s},t}(X_\rr{s})\Phi_{\rr{f},t}(X_{\rr{s}})^{-1}$,
where by definition $X_\rr{f}=\phi_{\rr{f},t}(X_\rr{s})$.

Let $J_{\alpha,t}:=|F_{\alpha,t}|$, with $\alpha=\rr{s},\rr{f}$, 
be the Jacobian of the transformation 
$\chi_{\alpha,t}$ measuring the ratio between current and 
reference volumes; we define the Green--Lagrange strain of the solid
$\varepsilon:=(F^{\top}_{\rr{s},t}F_{\rr{s},t}-I)/2$, where $I$ is the second order identity tensor.

From now on we shall restrict our attention just to one--dimensional space of positions, having in mind to consider applications of the present model to the so--called consolidation problem \citep[see][]{biot41,Terzaghi46,Cryer63}. In this framework tensorial quantities restrict to scalars; in particular the Green--Lagrange strain reduces to $\varepsilon:=(J_{\rr{s},t}^2-1)/2$.\medskip

\par\noindent
\textit{Mass balance.\/}
Let $\varrho_{0, \alpha}:\mathcal B_\alpha\to I\!\!R$, with $\alpha=\rr{s},\rr{f}$,
be the solid and fluid reference \textit{densities}.
The \textit{total mass} of each of the two components,
\begin{equation}
M_\alpha:=\int_{\mathcal B_\alpha}\varrho_{0,\alpha}(X_\alpha)\,\rr{d}X_{\alpha}
\end{equation}
with $\alpha=\rr{s},\rr{f}$, is supposed to be constant with respect to time 
$t$, which means that the mass is conserved when passing from the reference to the current configuration. Let $\varrho_{\alpha,t}$, with $\alpha=\rr{s},\rr{f}$, the solid and fluid current densities, mass conservation reads
\begin{equation}
\label{global_mass}
\dint_{\mathcal B_\alpha}\varrho_{0,\alpha}(X_{\alpha})\,\rr{d}X_{\alpha}
=\dint_{\mathcal B}\varrho_{\alpha,t}(x)\,\rr{d}x \medskip
=\dint_{\mathcal B_\alpha}\,\varrho_{\alpha,t}(\chi_{\alpha,t}(X_{\alpha}))\,
J_{\alpha,t}(X_{\alpha})\rr{d}X_{\alpha}
\end{equation}
which in the local form, i.e. $\forall X_{\alpha}\in \mathcal B_\alpha$, becomes
$\varrho_{\alpha,t}(\chi_{\alpha,t}(X_{\alpha}))J_{\alpha,t}(X_{\alpha})=
  \varrho_{0,\alpha}(X_{\alpha})$. Note that the initial value $\varrho_{\alpha,0}$ of the current density of the $\alpha$-th constituent can be assumed equal to the reference density $\varrho_{0, \alpha}$. Using the map $\phi_{\rr{f},t}$ allows for introducing the solid Lagrangean mass density of the fluid constituent: 
\begin{equation}
m_{\rr{f},t}(X_{\rr {s}}):=\varrho_{0,\rr{f}}(\phi_{\rr{f},t}(X_{\rr{s}}))\rr{det}\Phi_{\rr{f},t}(X_{\rr{s}})
\end{equation}

\noindent The admissible deformations of the porous continuum are therefore completely known once the Green--Lagrange strain and the solid Lagrangean mass density of the fluid are determined.\medskip

\par\noindent
\textit{Overall potential.\/}
To study the equilibrium property of the system at constant temperature a suitable overall potential energy $\Phi$, per unit volume, given by the sum of the \textit{Helmoltz free energy} $\Psi$ and the potential of external forces, can be introduced. Since our goal is that of modeling soil consolidation, external loading will only consist of a pure pressure acting on the solid skeleton, which implies the overall potential to be $\Phi=\Psi+pJ_{\rr{s}}$. 

As already noticed in a one--dimensional space of positions $\varepsilon=(J_{\rr{s}}^2-1)/2$,
if we restrict the discussion to the regime of small deformations, 
namely $J_{\rr{s}}\approx1$ and $\varepsilon\approx0$, we can expand around $J_{\rr{s}}=1$ and 
get $\varepsilon\approx J_{\rr{s}}$ (\textit{geometrical linearization}). 

For the sake of simplicity, from now on we shall denote with $m$ the increment of fluid mass density $m_{\rr f}$, with respect to a suitable reference value $m_{0,\rr f}$. For small deformations a reasonable expression for the dimensionless potential density of isotropic porous materials is given, in the framework of the \citet{biot41,biot55} theory, by the following quadratic form
\begin{equation}
\label{biot}
\Phi^\rr{B}(m,\varepsilon)=
p\varepsilon+\frac{1}{2}\varepsilon^2+\frac{1}{2}a(m-b\varepsilon)^2
\end{equation}
where $a>0$ is the ratio between the Biot modulus $M$ \citep[see][]{Coussy_book04}, and the solid bulk modulus, while $b>0$ is the so--called Biot coefficient and measures the coupling between the solid and the fluid components; $p$ is made dimensionless with respect to the solid bulk modulus. It is immediate to show that the only stationary state of 
(\ref{biot}) is $m_\rr{B}(p)=-bp$ and 
$\varepsilon_\rr{B}(p)=-p$.
Our hope is to describe the phase transition, driven by the pressure $p$,
by considering additional third and fourth order terms in the overall potential.
In this perspective we introduce a fourth order potential which reduces to the Biot 
one for small values of the parameters $m$ and $\varepsilon$, namely,
in our expression, the second order terms will be precisely the Biot 
ones. We set
\begin{equation}
\label{gibbs}
\Phi(m,\varepsilon)=
\frac{1}{12}\alpha m^2(3m^2-8b\varepsilon m+6b^2\varepsilon^2)+
\Phi^{\rr{B}}(m,\varepsilon)
\end{equation}
where $\alpha=\alpha(a,b)$ is a positive real function 
of the physical parameters $a$ and $b$.
While the function $\alpha$ can be chosen freely to tune the results
with reasonable physical behaviors,
the coefficients of the second order 
trinomial have been chosen so that (\ref{gibbs}) admits the local
minimum $(m_1(p),\varepsilon_1(p))$, with $m_1=b\varepsilon_1$,
describing the \textit{trivial} phase of the system similar to the one obtained
in the framework of the Biot theory.

\section{P\lowercase{hase transition}}
\label{s:transizione}
\par\noindent
We discuss the stationary states of the system which are 
identified with the minima of the two variables 
function $\Phi(\varepsilon,m)$.
We shall show that a point similar to the Biot one always exists, while,
depending on the pressure, more precisely for a sufficiently large 
pressure, a second stationary state shows up.
This section is devoted to the mathematical discussion of the 
phenomenon, its physical interpretation is postponed to 
the Section~\ref{s:risultati}.

\subsection{Stationary points of the overall potential}
\label{s:esistenza}
In order to find the stationary points of the overall potential
we compute the first order partial derivatives of the potential energy $\Phi$
\begin{equation}
\label{fasi00a}
\Phi_{m}=
(m-b\varepsilon)(\alpha m^2-\alpha b\varepsilon m+a)
\end{equation}
and
\begin{equation}
\label{fasi00b}
\Phi_{\varepsilon}=
\frac{1}{3}\alpha bm^2(3b\varepsilon-2m)
 -ab(m-b\varepsilon)
 +\varepsilon+p
\end{equation}
Letting $\Phi_{m}=0$ we get the following three solutions: 
$m_1=b\varepsilon$ and 
\begin{equation}
\label{fasi01}
m_\pm=
\frac{b}{2}
\Big[
     \varepsilon\pm\sqrt{\varepsilon^2-\frac{4a}{\alpha b^2}}
\Big]
\end{equation}

To get the corresponding values of $\varepsilon$, the equation 
$\Phi_{\varepsilon}=0$ must be solved with $m=m_1,m_\pm$.
We first note that by inserting $m=m_1$ in 
$\Phi_{\varepsilon}=0$, we get 
\begin{equation}
\label{fasi02}
p=-\varepsilon-\frac{1}{3}\alpha b^4\varepsilon^3
=:f_1(\varepsilon)
\end{equation}
Studying the function $f_1$ it is immediate to deduce that (\ref{fasi02}) 
has a single real solution for any $p>0$; we let $\varepsilon_1$ such a unique 
solution and remark that the stationary point $(m_1,\varepsilon_1)$ of 
the overall potential does 
exist for any choice of the parameter of the model.

To solve (\ref{fasi00b}) with $m=m_\pm$, we bound the discussion 
to the physically relevant region $\varepsilon<0$. Note that, if $|\varepsilon|\geq 2/(b\sqrt{\alpha/a})$ then
\begin{equation}
\label{fasi03}
m_\pm=
\frac{1}{2}b\varepsilon
\Big[
     1\mp\sqrt{1-\frac{4a}{\alpha b^2\varepsilon^2}}
\Big]
<0
\end{equation}
which implies, in particular, that $m_+\geq m_->m_1$;
remark that for $|\varepsilon|$ large enough, $m_+$ approaches $0$ and 
$m_-$ approaches $m_1$.
By letting $m=m_\pm$ in (\ref{fasi00b}), we get 
\begin{equation}
\label{fasi04}
p
=
-\varepsilon+ab[m_\pm(\varepsilon)-b\varepsilon]
  -\alpha b^2\varepsilon m_\pm^2(\varepsilon)
  \vphantom{\bigg\{}
  +\frac{2}{3}\alpha bm_\pm^3(\varepsilon)
  =:f_\pm(\varepsilon)
\end{equation}
Since $f_\pm$ are not defined around $\varepsilon=0$, namely, for 
$|\varepsilon|< 2/(b\sqrt{\alpha/a})$, we can hope to describe 
some phase transition driven by $p$; indeed we guess that for 
$p$ small enough the equation (\ref{fasi04}) will have no real solution
so that $(m_1,\varepsilon_1)$ will be the sole stationary point
of the overall potential.
\begin{figure}
 \centering
 \includegraphics[width=9cm]{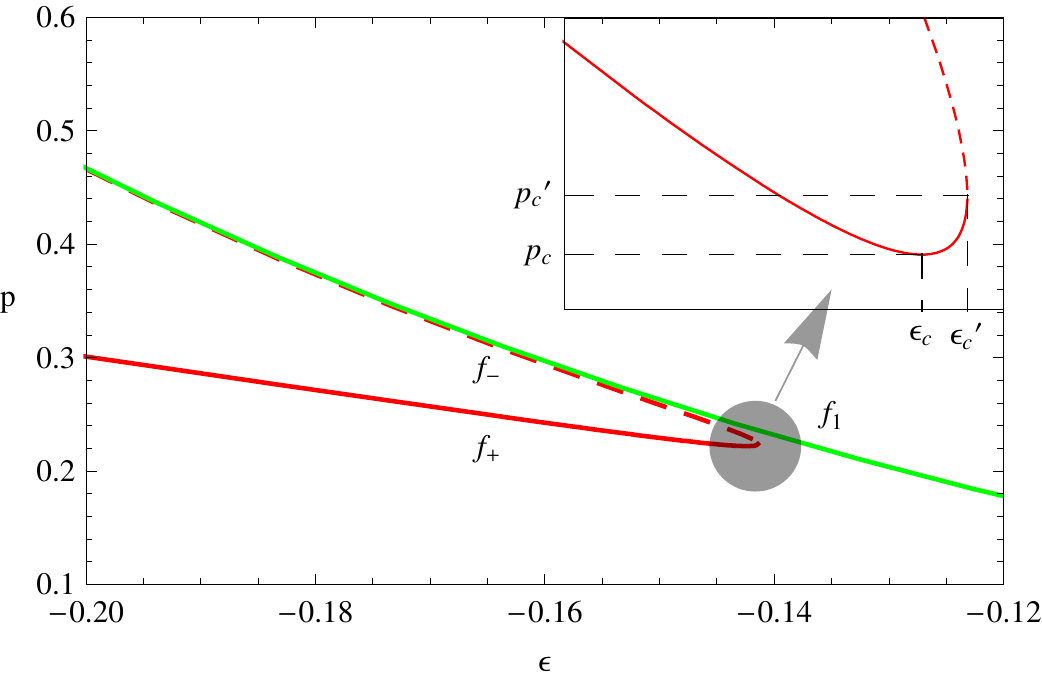}
 % fpmf1.pdf: 300x195 pixel, 72dpi, 10.58x6.88 cm, bb=0 0 300 195
\caption{The graph of $f_{\pm}$ and $f_1$ is plotted. Note that the dashed line corresponds to the graph of $f_-$. The values of these functions at $p$ determine the states of equilibrium of the porous medium. The used values of the constitutive parameters are $a=1/2$, $b=1$, $\alpha=100$.}
\label{f1fpm}
\end{figure}
To explain our guess we draw, 
in Figure \ref{f1fpm},
$f_1,f_\pm$ as function of $\varepsilon$ for the 
parameters specified in the captions; note that 
$f_-$ approaches $f_1$ for $|\varepsilon|$ large, this is consistent 
with the behavior of $m_1$ and $m_-$.
We let $p_\rr{c}$ be the minimum of the function $f_+$ and 
$p'_\rr{c}:=f_+(\varepsilon'_\rr{c})=f_-(\varepsilon'_\rr{c})$
with $\varepsilon'_\rr{c}:=-2/(b\sqrt{\alpha/a})$, the largest value of $\varepsilon$ for which the additional stationary point of $\Phi$ appears.
For $p<p_\rr{c}$ the solutions $(m_\pm,\varepsilon_\pm)$ are not real,
hence the system has a single phase, the one essentially due to the Biot model. 
For $p_\rr{c}<p<p'_\rr{c}$, the solution $(m_+,\varepsilon_+)$ 
with the good $p$--behavior (the one with smallest $\varepsilon_+$)
should be interpreted as the second phase, while 
the other should be a saddle point of the overall potential.
For $p>p'_\rr{c}$, 
the unique solution $(m_+,\varepsilon_+)$ 
should be interpreteted as the second phase, while 
$(m_-,\varepsilon_-)$ should be a saddle point of the overall potential.
To prove that this interpretation is correct one should study 
the second order derivatives of $\Phi$ (see \S \ref{s:carattere}).

To study the equation (\ref{fasi04}) we recall that 
the two functions $f_\pm$ are defined for 
$\varepsilon<\varepsilon'_\rr{c}:=-2/(b\sqrt{\alpha/a})$
and note that 
\begin{equation}
\label{pcritica}
p'_\rr{c}
:=
f_\pm(\varepsilon'_\rr{c})
=
\frac{1}{b\sqrt{\alpha/a}}
\Big(
2+\frac{7}{3}b^2a
\Big)
>0
\end{equation}

To study the $\varepsilon\to-\infty$ limit, we recall that 
$\sqrt{1-x}=1-x/2-x^2/8+O(x^3)$, and using (\ref{fasi03}) we get 
\begin{equation}
m_\pm(\varepsilon)=\frac{1}{2}b\varepsilon
\Big[
     2\delta_{-,\pm}
     \pm\frac{2a}{\alpha b^2\varepsilon^2}
     \pm\frac{2a^2}{\alpha^2 b^4\varepsilon^4}
     +O(1/\varepsilon^6)
\Big]
\end{equation}
where $\delta_{-,\pm}$ is the Kronecker delta, such that $\delta_{-,-}=1$ and $\delta_{-,+}=0$. Substituting those 
expansions in the definition (\ref{fasi04}) of $f_\pm$ and keeping
track of the not vanishing terms we get
\begin{equation}
\label{comportamento}
f_+(\varepsilon)\sim-\varepsilon(1+ab^2)
\textrm{ and }
f_-(\varepsilon)\sim-\varepsilon-\frac{1}{3}\alpha b^4\varepsilon^3
\end{equation}
Note that for $\varepsilon\to-\infty$ the function $f_-$ behaves precisely 
as $f_1$. 

We now compute the first order derivative. 
Recalling the definition (\ref{fasi04}) of $f_\pm$, see equation (\ref{fasi00a}), and the 
fact that $m_\pm$ are solutions of the equation $\Phi_{m}=0$, we
get 
\begin{equation}
\label{derivatafpm}
\begin{array}{rrl}
f'_\pm(\varepsilon)
& \!\!=\!\!  &
-1-ab^2-\alpha b^2m_\pm^2(\varepsilon)
+bm_\pm'(\varepsilon)
 [a-2\alpha b \varepsilon m_\pm(\varepsilon)+2\alpha m_\pm^2(\varepsilon)]
\\
& \!\!=\!\!  &
-1-ab^2-\alpha b^2m_\pm^2(\varepsilon)-abm_\pm'(\varepsilon)
\\
\end{array}
\end{equation}
Note the the first three terms are clearly negative, so that the sign 
of the derivatives depends essentially on the sign of the fourth term. We 
then recall (\ref{fasi03}), which is valid for $\varepsilon<0$, and compute 
\begin{equation}
\label{derivataemme}
m_\pm'(\varepsilon)=
\frac{1}{\varepsilon}m_\pm(\varepsilon)
   \mp\frac{2a}{\alpha b\varepsilon^2}
   \Big[
        1-\frac{4a}{\alpha b^2\varepsilon^2} 
   \Big]^{-1/2}
\end{equation}
By using (\ref{derivataemme}) and (\ref{derivatafpm}) it follows 
immediately that $f'_\pm(\varepsilon)\to\pm\infty$ for 
$\varepsilon\to\varepsilon'_\rr{c}$ from the left.

Notice that $m_-'(\varepsilon)$, for $\varepsilon<0$, is positive. Hence, 
by (\ref{derivatafpm}) it follows that $f_-'(\varepsilon)<0$ for 
$\varepsilon<\varepsilon'_\rr{c}$.
The function $f_-$ decreases monotonously from $+\infty$ to $p'_\rr{c}$ 
as $\varepsilon$ goes from $-\infty$ to $\varepsilon'_\rr{c}$.
We then conclude that the stationary point $(m_-,\varepsilon_-)$ of the 
overall potential exists and is unique for $p\ge p'_\rr{c}$. 

The study of $f_+$ is slightly more difficult. 
First of all we recall that $0>m_+(\varepsilon)>b\varepsilon$, see
(\ref{fasi03}), hence, using the definition (\ref{fasi04}), we get 
\begin{equation}
f_+(\varepsilon)\!=\!
-\varepsilon+ab(m_+(\varepsilon)-b\varepsilon)
+\alpha bm^2_+(\varepsilon)\Big[\frac{2}{3}m_+(\varepsilon)-b\varepsilon\Big]
\!>0
\end{equation}
Thus $f_+(\varepsilon)$ is a positive function tending to
$+\infty$ for $\varepsilon\to-\infty$ and approaching 
$f_+(\varepsilon'_\rr{c})>0$, see (\ref{pcritica}), with 
positive slope. Hence there must be at least a minimum of the 
function $f_+(\varepsilon)$ in the region 
$\varepsilon\in(-\infty,\varepsilon'_\rr{c}]$.
Moreover, it is possible easy to see that the equation $f'_+(\varepsilon)=0$ 
is biquadratic in $\varepsilon$. Hence it can have either zero or one or two negative solutions, the only possible case, compatible with the above mentioned properties of $f_+$, is that the negative solution is unique.

Finally, we set $p_\rr{c}:=f_+(\varepsilon_\rr{c})>0$ and remark that 
for $p<p_\rr{c}$ the equations $f_\pm(\varepsilon)=p$ have no 
real solution.
For $p_\rr{c}<p<p'_\rr{c}$, 
the equation $f_-(\varepsilon)=p$ has no real solution, while
$f_+(\varepsilon)=p$ has two real solutions
$(m_+^1,\varepsilon_+^1)$ and $(m_+^2,\varepsilon_+^2)$ 
with $\varepsilon_+^1<\varepsilon_+^2$.
For $p>p'_\rr{c}$, 
both $f_-(\varepsilon)=p$ and $f_+(\varepsilon)=p$ have a unique 
real solution, respectively denoted by
$(m_-,\varepsilon_-)$ and $(m_+^1,\varepsilon_+^1)$.
According to the above depicted scenario, see the discussion 
below (\ref{fasi04}), we expect that 
$(m_+^1,\varepsilon_+^1)$ is a minimum of the overall potential,
while $(m_+^2,\varepsilon_+^2)$ and $(m_-,\varepsilon_-)$ are 
saddle points. The validity of this guess will be proven in 
the next section.

\subsection{Character of the stationary points}
\label{s:carattere}
We study, now, the character of the stationary points of the overall potential.
Computing the second order derivatives of $\Phi$ with respect to $m$ and 
$\varepsilon$ we get
\begin{equation}
\label{hess01}
\begin{array}{rcl}
{\displaystyle
 \Phi_{mm}
}
& =  &
(\alpha m^2-\alpha b\varepsilon m+a)
+(m-b\varepsilon)(2\alpha m-\alpha b\varepsilon)
\vphantom{\Big\{}
\\
{\displaystyle
 \Phi_{m\varepsilon}
}
& =  &
-ab+2\alpha bm(b\varepsilon-m)
\\
{\displaystyle
 \Phi_{\varepsilon\varepsilon}
\vphantom{\Big\{}
}
& = &
1+ab^2+\alpha b^2m^2>0
\vphantom{\Big\{}
\\
\end{array} 
\end{equation}

Using that $m_1=b\varepsilon_1$, we are able to compute the Hessian 
$\cc{H}(m_1,\varepsilon_1)=a(1+\alpha m_1^2 b^2)>0$ and, thus, 
conclude that $(m_1,\varepsilon_1)$ is a local minimum of 
(\ref{gibbs}). Hence it represents a stationary state 
of the model.

We have to study, now, the properties of the 
stationary points $(m_\pm,\varepsilon_\pm)$.
In view of this we 
give a nice expression of the Hessian computed in 
$(m_\pm(\varepsilon),\varepsilon)$.
By using (\ref{hess01}) and recalling that $m_\pm(\varepsilon)$
is obtained by solving the 
equation $\Phi_{m}=0$, see (\ref{fasi00a}), we get 
\begin{equation}
\label{hess02}
\begin{array}{rrl}
{\displaystyle
 \Phi_{mm}
 (m_\pm(\varepsilon),\varepsilon)
}
& =  &
\alpha (m_\pm(\varepsilon)-b\varepsilon)(2m_\pm(\varepsilon)-b\varepsilon)
\vphantom{\Big\{}
\\
{\displaystyle
 \Phi_{m\varepsilon}
 (m_\pm(\varepsilon),\varepsilon)
}
& = &
ab
\vphantom{\Big\{}
\\
{\displaystyle
 \Phi_{\varepsilon\varepsilon}
 (m_\pm(\varepsilon),\varepsilon)
}
& \!\!=\!\!  &
1+b^2(a+\alpha (m_\pm(\varepsilon))^2)>0
\vphantom{\Big\{}
\\
\end{array} 
\end{equation}
so that 
\begin{equation}
\begin{array}{rrl}
\cc{H}_\pm(\varepsilon)
:=
\cc{H}(m_\pm(\varepsilon),\varepsilon)
\vphantom{\Big\{}
& =  &
[1+b^2(a+\alpha(m_\pm(\varepsilon))^2)]
[\alpha (m_\pm(\varepsilon)-b\varepsilon)(2m_\pm(\varepsilon)-b\varepsilon)]
-a^2b^2
\vphantom{\Big\{}
\\
& = &
[1+b^2\alpha b\varepsilon m_\pm(\varepsilon))]
[\alpha (m_\pm(\varepsilon)-b\varepsilon)(2m_\pm(\varepsilon)-b\varepsilon)]
-a^2b^2
\\
\end{array}
\end{equation}
By using (\ref{fasi01}) we get 
\begin{equation}
\cc{H}_\pm(\varepsilon)
=
-2a-a^2b^2+\alpha b^2\varepsilon^2/2
\mp\alpha b^2(1/2+ab^2)\varepsilon\sqrt{\varepsilon^2
   -4a/(\alpha b^2)}
\end{equation}
Since we have focused our discussion on the case $\varepsilon<0$, 
we have that 
\begin{equation}
\label{hess05}
\cc{H}_\pm(\varepsilon)
=
-2a-a^2b^2+\alpha b^2\varepsilon^2/2
\pm\alpha b^2(1/2+ab^2)\varepsilon^2\sqrt{1
   -4a/(\alpha b^2\varepsilon^2)}
\end{equation}

We use the above expression (\ref{hess05}) to prove that the stationary 
point $(m_-,\varepsilon_-)$ is a saddle point. First we note that 
if such a stationary point exists then it must be necessarily
$\varepsilon_-\le\varepsilon'_\rr{c}$, 
so that $4a/(\alpha b^2\varepsilon_-^2)\le1$.
Remarking that, for $0\le x\le1$, one has $\sqrt{1-x}>1-x$, by (\ref{hess05}) 
we get the bound 
\begin{equation}
\begin{array}{rrl}
\cc{H}_-(\varepsilon_-)
& \le  &
-2a-a^2b^2+\alpha b^2\varepsilon_-^2/2
\vphantom{\Big\{}
-\alpha b^2(1/2+ab^2)\varepsilon_-^2
   [1-4a/(\alpha b^2\varepsilon_-^2)]
\vphantom{\Big\{}
\\
& =  &
ab^2[3a-\alpha b^2\varepsilon_-^2]\le a b^2\left(3a -4a\right)=-a^2b^2<0
\vphantom{\Big\{}
\\
\end{array}
\end{equation}
Where the last bound follows from the fact that, since 
$\varepsilon_-\le\varepsilon'_\rr{c}$, we have 
$\alpha b^2\varepsilon_-^2\ge4a$.

In order to study the character of the stationary points 
$(m_+^i,\varepsilon_+^i)$, with $i=1,2$, we study the function 
$\cc{H}_+(\varepsilon)$ for $\varepsilon<0$.
Recall (\ref{hess05}) and note that
$\cc{H}_+(\varepsilon'_\rr{c})=-a^2b^2<0$
and $\cc{H}_+(\varepsilon)\to+\infty$ for $\varepsilon\to-\infty$.
By computing the first derivative of $\cc{H}_+(\varepsilon)$ with respect
to $\varepsilon$ we get
\begin{equation}
\label{hess06}
\cc{H}'_+(\varepsilon)
=
\alpha b^2\varepsilon
\Big[
     1
      +\Big(\frac{1}{2}+ab^2\Big)
            \Big(1-\frac{4a}{\alpha b^2\varepsilon^2}\Big)^{-1/2}
\Big(2-\frac{4a}{\alpha b^2 \varepsilon^2}\Big)\Big]
\end{equation}
which is clearly negative for $\varepsilon<\varepsilon'_\rr{c}$. 
Thus, we have that 
$\cc{H}_+(\varepsilon)$ decreases from $+\infty$ to $-a^2b^2$ as 
$\varepsilon$ goes from $-\infty$ to $\varepsilon'_\rr{c}$. 
Denoted by $\bar\varepsilon$ the unique negative zero of 
$\cc{H}_+(\varepsilon)$, 
we prove that $\bar\varepsilon=\varepsilon_\rr{c}$, that is 
$\bar\varepsilon$ is the minimum of the function $f_+(\varepsilon)$. 
Note, indeed, that by definition 
\begin{equation}
\label{hess10}
\cc{H}_+(\varepsilon)= 
\Phi_{mm}(m_+(\varepsilon),\varepsilon)
\Phi_{\varepsilon\varepsilon}(m_+(\varepsilon),\varepsilon)
-
\Phi^2_{m\varepsilon}(m_+(\varepsilon),\varepsilon)
\end{equation}
Recall that the function $m_+(\varepsilon)$ is implicitely defined by 
the equation 
$\Phi_{m}(m_+(\varepsilon),\varepsilon)=0$
and note that (\ref{fasi04}) is equivalent to 
$f_+(\varepsilon)
 =-\Phi_{\varepsilon}(m_+(\varepsilon),\varepsilon)+p$.
By using the chain rule and the implicit function theorem
we get that 
\begin{equation}
\begin{array}{rrl}
{\displaystyle
 f'_+(\varepsilon)
}
&=&
{\displaystyle
 -\Phi_{\varepsilon\varepsilon}(m_+(\varepsilon),\varepsilon)
 -\Phi_{\varepsilon m}
                              (m_+(\varepsilon),\varepsilon)
			      \,m'_+(\varepsilon)
=
 -\Phi_{\varepsilon\varepsilon}(m_+(\varepsilon),\varepsilon)
 +\frac{\Phi^2_{\varepsilon m}(m_+(\varepsilon),\varepsilon)}
       {\Phi_{mm}(m_+(\varepsilon),\varepsilon)}
 \vphantom{\Big\{}
}
\\
&=&
{\displaystyle
 -\frac{1}{\Phi_{mm}(m_+(\varepsilon),\varepsilon)}
  \,\cc{H}_+(\varepsilon)
 \vphantom{\Big\{}
}
\\
\end{array}
\end{equation}
where in the last step we have used (\ref{hess10}).

The above equality ensures that $\bar\varepsilon=\varepsilon_\rr{c}$. 
Finally, since, 
$\varepsilon^1_+<\varepsilon_\rr{c}<\varepsilon^2_+$, we have that 
$(m^1_+,\varepsilon^1_+)$ is a minimum of (\ref{gibbs}) 
while 
$(m^2_+,\varepsilon^2_+)$ is a saddle.

\section{R\lowercase{esults}}
\label{s:risultati}
In this section a discussion on the behavior of the minima of the overall potential (\ref{gibbs}), when varying the external pressure, will be presented. A parametric analysis in dependence of the coefficients introduced in the Biot model, $a$ and $b$, as well as of the additional coefficient $\alpha\left(a,b\right)$ multiplying the fourth order terms of (\ref{gibbs}) will be also developed. 

According to the experimental results on fluidization of soils, cited in the introduction \citep[see][]{Nichols94, Holcomb03, Holcomb07, Vardoulakis04_1, Vardoulakis04_2}, the proposed model is capable to exhibit, in the presence of an applied consolidating external pressure, an additional stationary state in the $\left(\varepsilon,m\right)$ plane beyond the one corresponding to classical consolidation. Increasing of the external pressure, acting on the porous medium, induces, on one hand, part of the fluid to flow out of the skeleton, on the other, part of the fluid to be segregated into not connected cavities of the solid matrix, when the external pressure overwhelms a critical value.

\begin{figure}[htp]
 \centering
 \includegraphics[width=9cm]{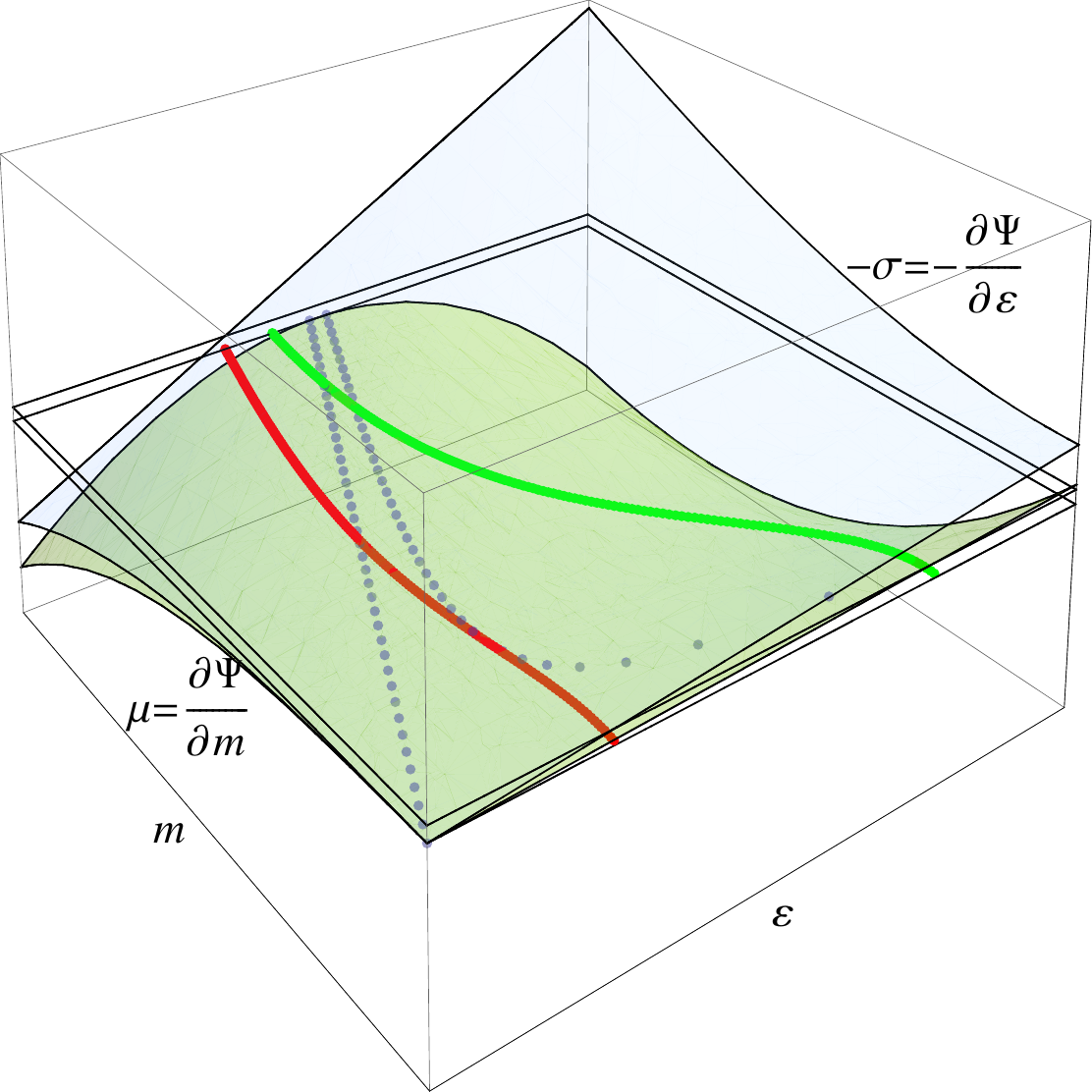}
 % FigEquilib1.pdf: 320x283 pixel, 72dpi, 11.29x9.98 cm, bb=0 0 320 283
\caption{Cutting the $-\sigma$ surface with the horizontal plane at $p$ identifies the root locus of $\Phi_{\varepsilon}=0$; in particular the two solid lines identify the root loci associated with the critical value of the applied pressure and that associated to a larger value of $p$. The dotted lines describe the root locus of $\Phi_{m}=0$, obtained cutting the $\mu$ surface with the $\left(\varepsilon,m\right)$ plane.} 
\label{equilibria}
\end{figure}

\begin{figure}[htp]
 \centering
 \includegraphics[width=9cm]{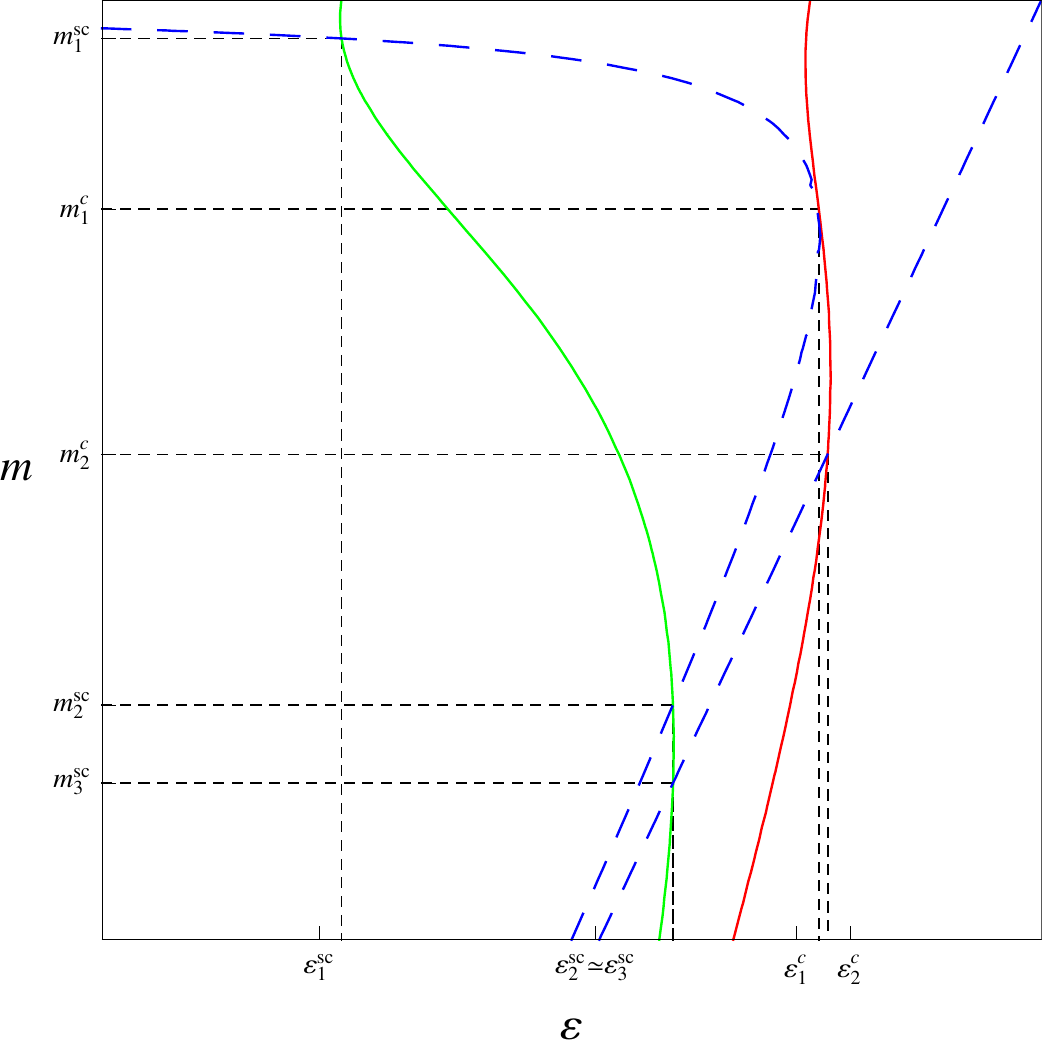}
 % Proiezione.pdf: 300x297 pixel, 72dpi, 10.58x10.48 cm, bb=0 0 300 297
\caption{This is a detail of the projection on the $\left(\varepsilon,m\right)$ plane of the intersection of $-\sigma$ and $\mu$ with the horizzontal plane at $p$ and zero, respectively. In particular $\left(\varepsilon_i^c,m_i^c\right)$, $i=1,2$, identify the two stationary points of (\ref{gibbs}) when $p=p_c$, while $\left(\varepsilon_i^{sc},m_i^{sc}\right)$, $i=1,2,3$, identify the three stationary points of (\ref{gibbs}) when $p>p_c$ (super--critical conditions).} 
\label{proiez_equilibria}
\end{figure}

The arising of the additional state of equilibrium, associated with pore--fluid segregation, is explicitly depicted in Fig. (\ref{equilibria}), where the opposite of the total stress $\sigma=\partial \Psi/\partial\varepsilon$ and the fluid chemical potential $\mu=\partial \Psi/\partial m$ are plotted against the kinematical parameters $\varepsilon$ and $m$. The projection over the $\left(\varepsilon,m\right)$ plane of the curves obtained cutting the $-\sigma$ surface with the horizontal plane at $p$ and the $\mu$ surface with the $\left(\varepsilon,m\right)$ plane itself have more than one mutual intersection, each of them identifies one of the stationary points of the overall potential (\ref{gibbs}). In particular the dotted (dashed) lines in Fig. (\ref{equilibria}) (Fig. (\ref{proiez_equilibria})) correspond to the solutions of $\Phi_m=0$, on the other hand the solid ones correspond to the solution of $\Phi_{\varepsilon}=0$, for different values of $p$.

%AGGIUNGERE NELLA CAPTION DETTAGLIO DEI PUNTI RAPPRESENTATI
As already noticed, the critical pressure $p_c$ is the smallest value of the consolidating loading for which the pore--fluid segregation solution shows up.

\begin{figure}
 \centering
 \includegraphics[width=9cm]{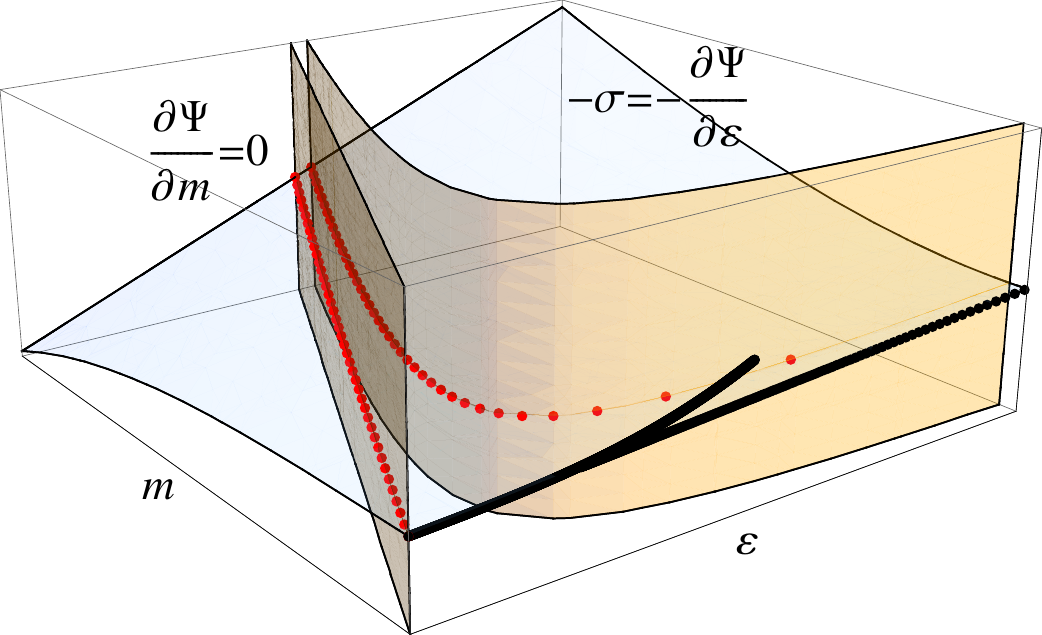}
 % VdWf1fp.pdf: 320x172 pixel, 72dpi, 11.29x6.07 cm, bb=0 0 320 172
\caption{The dotted lines correspond to the intersection between $-\sigma$ and the root locus of $\Phi_m=0$, the solid lines are the projections of these curves on the stress--strain plane, i.e. the $f_1$ and $f_{\pm}$ curves.}
\label{VdW_analog}
\end{figure}

With another point of view, the stationarity conditions stated by eq. (\ref{fasi00a}) and (\ref{fasi00b}) describe two different surfaces in the $(\varepsilon,m)$ plane, the intersection of which gives rise to a curve in the three--dimensional space, the dotted curve in Fig. (\ref{VdW_analog}), playing the role of the curve of the fluid pressure in terms of the specific volume, in the case of Van der Waals' model of liquid--vapor coexistence. It is worth to notice that the projection of these curves on the $\left(-\sigma,\varepsilon\right)$ plane provides the plot of the $f_1$ and $f_{\pm}$ functions, depicted in Fig. (\ref{f1fpm}). As in the case of Van der Waals' model sectioning this curve in the three dimensional space with an horizzontal plane at $p$ identifies one stationary (equilibrium) point, if $p<p_c$, or three stationary points, if $p\geq p_c$, two of which corresponds to classical consolidation and pore--fluid segregation solutions, respectively.

\begin{figure}[htp]
 \centering
 \includegraphics[width=16cm]{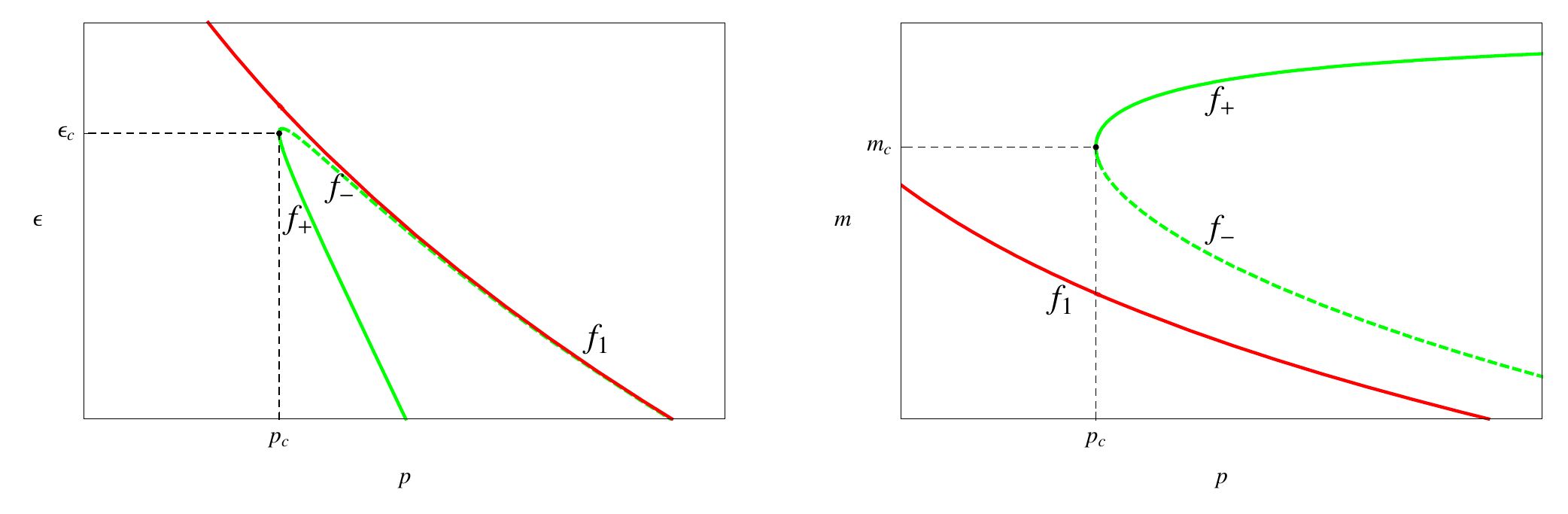}
 % Plotmp.pdf: 300x197 pixel, 72dpi, 10.58x6.95 cm, bb=0 0 300 197
\caption{The graph of $f_1$ corresponds to the classical behavior of consolidating soils, the negative strain progressively increases and the mass of the fluid progressively decreases when the external pressure raises up; the graph of $f_+$, on the other hand, describes the phenomenon of fluid--segregation, the dashed line (graph of $f_-$) corresponds to the saddle point of the overall potential.} 
\label{fig_emp}
\end{figure}

Fig. (\ref{fig_emp}) explicitly shows the parametrization with pressure of the states of equilibrium: when the applied external pressure reaches the critical value $p_c$ a new state of equilibrium appears in which the variation of the fluid mass and the negative strain increase. This is what is generally called fluid--segregation: under the applied external pressure the porous material undergoes to consolidation, however part of the fluid remains trapped into non--connected cavities of the skeleton, the formation of which is possibly due to the thinnering of duct connections of the pore--space. 

\begin{figure}[htp]
 \centering
 \includegraphics[width=16cm]{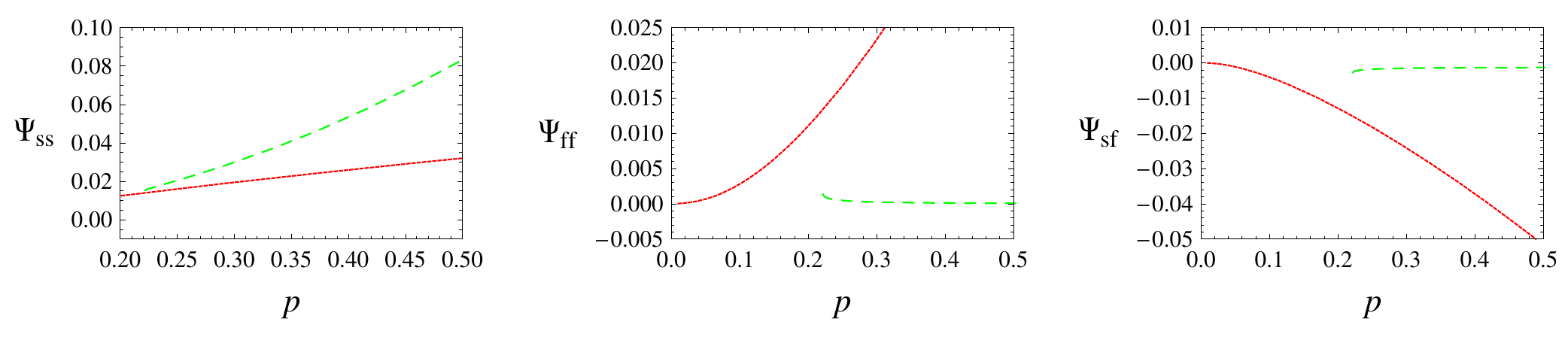}
 % EnergyPlot.pdf: 600x129 pixel, 72dpi, 21.17x4.55 cm, bb=0 0 600 129
 \caption{The solid lines indicate the \textit{pure solid}, the \textit{pure fluid} and the \textit{coupling} energies associated with the soil consolidating equilibrium path; the dashed ones indicate the same energy contributions associated with the pore--fluid segregation equilibrium.}
\label{EnergyPlot}
\end{figure}

Note that all the pictures have been drawn for the parameters specified in the caption of Fig.(\ref{f1fpm}). However it is important to stress that, according to the general discussion of \S \ref{s:transizione}, no variation in the main features of the considered phase transition, from the classical consolidating soil phase towards the pore--fluid segregated one, can arise when tuning the constitutive parameters $a$, $b$ and $\alpha$. The only constraint on the showing up of the new equilibrium being the impenetrability of the skeleton ($\varepsilon>-1/2$). It is worth to notice that the absolute values of both the free energy of the \textit{pure fluid} and that associated with solid--fluid coupling in the overall free energy (\ref{gibbs})
\begin{equation}\label{Psiff-Psisf}
\Psi_{\rr {ff}}\left( \varepsilon,m\right):=1/4\alpha m^4+1/2 a m^2\quad \rr{and} \quad \Psi_{\rr {sf}}\left( \varepsilon,m\right):=-2/3\alpha b m^3 \varepsilon+1/2\alpha b^2 m^2\varepsilon^2-a b m\varepsilon
\end{equation}
progressively decrease for increasing values of the applied external pressure, along the pore--fluid segregating equilibrium path, see Fig. (\ref{EnergyPlot}). Apparently this is due to the increasing incapability of the pore fluid to flow out of the skeleton. Conversely, the free energy of the skeleton, $\Psi_{\rr {ss}}:=1/2 \left(1+a b^2\right)\varepsilon^2$, along the pore--fluid segregating equilibrium is larger than the one associated to classical consolidation, the trapped fluid inducing a virtual stiffening of the solid matrix.

Phase transition is therefore completely characterized in terms of the dependence of the critical pressure on the ratio between the Biot modulus and the bulk modulus of the skeleton, $a$, and the Biot coefficient $b$. 
\begin{figure}[htp]
 \centering
 \includegraphics[width=9cm]{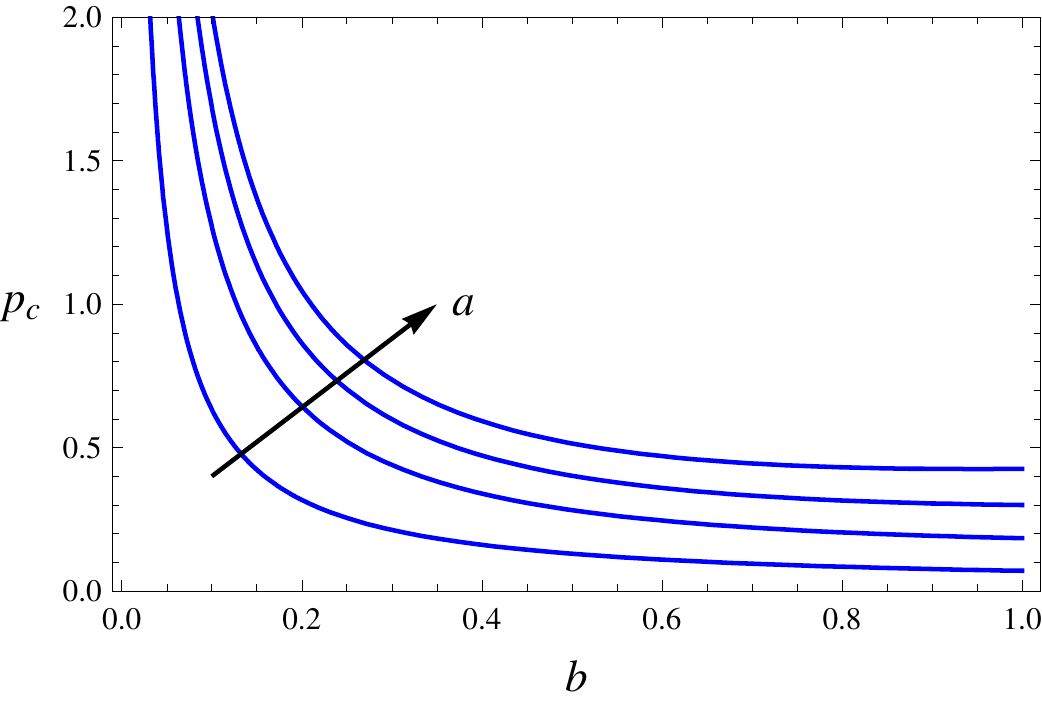}
 % Plotmp.pdf: 300x197 pixel, 72dpi, 10.58x6.95 cm, bb=0 0 300 197
 \caption{Picture of the variation of the critical pressure with the Biot coefficient $b$, parametrized by $a$; both $a$ and $b$ range in the open interval $\left(0,1\right)$.}
 \label{pcr_ab}
\end{figure}
In the frame of physically meaningful assumptions on these parameters ($a\leq 1$ and $b\leq 1$) the critical pressure $p_c$ exhibits the behavior depicted in Fig. (\ref{pcr_ab}). Increasing $a$, keeping $b$ fixed, implies the critical pressure to increase and therefore the phase transition to occur for higher values of the consolidating pressure. Indeed increasing $a$ means increasing of the bulk modulus of the fluid; the inverse of Biot's modulus is an affine function of the inverse of the fluid bulk modulus, with respect to that of the solid. The fluid becomes more stiff and therefore does not allow the applied external pressure to force shrinkage of the  ducts inside the porous medium; as a consequence flowing of the fluid out of the solid matrix is enhanced.

Increasing $b$, keeping $a$ fixed, on the other hand, implies the critical pressure to descrease and therefore the phase transition to occur for lower levels of the consolidating loading. This is due to the fact that increasing $b$ enhances solid--fluid coupling.

Apparently no physical reasoning can be developed in order to discuss the behavior of the system when tuning $\alpha$, however it is clear that only for sufficiently high values of this parameter phase transition can be appreciated.

%\bibliography{poro.bib}

\end{document}